\documentclass[12pt,a4paper]{article}
\usepackage{float}              
\usepackage{supertabular}
\usepackage{longtable}
\usepackage{epsfig}
\usepackage{rotating}
\usepackage{exscale}            
\usepackage{cite}               
\usepackage{graphicx}           
\usepackage{epstopdf}
\usepackage[rm,hang]{caption}   
\usepackage{amsfonts}           
\usepackage{amsmath}
\usepackage{amssymb}
\usepackage{ifthen}
\usepackage{xspace}
\usepackage{booktabs}
\usepackage[footnotesize]{subfigure}
\usepackage[dvips]{psfrag}
\usepackage{color,colortbl}
\usepackage{hyperref}

\usepackage[sc]{mathpazo}
\usepackage[T1]{fontenc}

\newcommand{\fref}[1]{Figure~\ref{#1}}
\newcommand{\tref}[1]{Table~\ref{#1}}

\renewcommand\arraystretch{1.5}

\renewcommand{\arraystretch}{0.5}

\newcommand{\eref}[1]{Equation~\eqref{#1}}

\renewcommand{\arraystretch}{0.65}

\makeatletter
\def\ps@paper{\def\@oddfoot{}\def\@oddhead{\hfil -- \thepage\ --\hfil}}
\begin{document}
\thispagestyle{empty}

\begin{center}
{\Large \bf 
Sampling Free Energy Surfaces as Slices by Combining Umbrella Sampling and Metadynamics
}\\
\vspace*{1.0cm}
{\bf \large  Shalini Awasthi, Venkat Kapil and Nisanth N. Nair$^\ast$\footnote{$^\ast$ Corresponding Author: nnair@iitk.ac.in }}
\\
\vspace*{0.5cm}
{\it Department of Chemistry \\
Indian Institute of Technology Kanpur, 208016 Kanpur, India }\\
\end{center}


%
%
\clearpage

%

 

\section*{Abstract}
Metadynamics (MTD) is a very powerful technique to sample high--dimensional free energy landscapes, and 
due to its  self--guiding property, the method has been successful in studying complex reactions and conformational
changes.
MTD sampling is based on filling the free energy basins by biasing potentials and thus for cases with 
flat, broad and unbound free energy wells, the computational time to sample 
them becomes very large.
To alleviate this problem, we combine the standard 
Umbrella Sampling (US) technique with MTD to sample orthogonal collective variables (CVs) in a simultaneous way.
Within this scheme, we construct the equilibrium distribution of CVs from biased distributions obtained from 
independent MTD simulations with umbrella potentials.
Reweighting is carried out by a procedure that combines US reweighting and Tiwary--Parrinello MTD reweighting within the
Weighted Histogram Analysis Method (WHAM).
The approach is ideal for a controlled sampling of a CV in a MTD simulation, making it computationally efficient
in sampling flat, broad and unbound free energy surfaces. 
This technique also allows for a distributed sampling of a high--dimensional free energy surface, further increasing 
the computational efficiency in sampling.
We demonstrate the application of this technique in sampling high--dimensional surface for various chemical reactions 
using
{\em ab initio} and QM/MM hybrid 
molecular dynamics simulations.
%
Further, in order to carry out MTD bias reweighting for computing forward reaction barriers in {\em ab initio } or QM/MM simulations, we propose a computationally affordable approach that does not require recrossing trajectories.
\clearpage

\section{Introduction}
Metadynamics (MTD)  is a very powerful  tool to sample complex free energy landscapes of complex chemical reactions, phase transitions and conformational changes.~\cite{Laio:PNAS:02,Iannuzzi:03,mtd:rev:11,Luigi:12,Gervasio:08}
MTD is a biased sampling approach where sampling of a selected set of collective variables (CVs) is enhanced 
by introducing slowly grown smooth history dependent repulsive potentials along the trajectory of CVs.
In MTD, the underlying potential energy of a system  $U(\mathbf R)$ is modified to
 $U(\mathbf R) + V^{\rm b}(\mathbf s,t)$, where $V^{\rm b}(\mathbf s,t)$ is the biasing potential applied
 along the CVs, $\mathbf s \equiv \mathbf s(\mathbf R)$, at any instance of the simulation $t$.
Typically, $V^{\rm b}(\mathbf s,t)$ is constructed by the sum of spherical Gaussians deposited over time, as
\begin{eqnarray}
 \label{eqn:b:mtd}
V^{\rm b}(\mathbf s,t) = \sum_{\tau < t} w_{\tau} \exp \left [ - \frac{ \left \{ \mathbf s - \mathbf s(\tau) \right \}^2 }{2 (\delta s)^2 } \right ]
\end{eqnarray}
Here $w_\tau$ and $\delta s$ are the height and the width parameters defining the Gaussian potential added at
some time $\tau$.
In Well--Tempered MTD (WT--MTD)~\cite{mtd:well:08},  
\begin{eqnarray}
w_{\tau}=\omega_0 \tau_{0} \exp \left [ - \frac{V^{\rm b}(\mathbf s, t)}{ k_{\rm B} \, \Delta T }\right ]
\end{eqnarray}
where $\omega_0$ is the initial rate of deposition of the bias, $\tau_{0}$ is the time step at which Gaussian potentials are augmented, 
and $\Delta T$ is a parameter.
The advantage of WT--MTD is that a systematic convergence in free energy can be achieved -- at the limit $t\rightarrow \infty$, the biasing potential $V^{\rm b}(\mathbf s,t)$ does not vary
much, and 
%
%
\begin{eqnarray}
 \lim_{t \rightarrow \infty} V^{\rm b}(\mathbf s,t) + f^\prime =  - \frac{ \Delta T}{T+\Delta T} F(\mathbf s)
\end{eqnarray}
where $f^\prime$ is a constant. Thus, a converged free energy surface can be constructed as
\begin{eqnarray}
\label{eqn:mtd:f}
F(\mathbf s) = - \alpha \lim_{t \rightarrow \infty}  V^{\rm b}(\mathbf s,t) + f 
\end{eqnarray}
where $\alpha = {(T + \Delta T)}/{\Delta T}$ and $f$ is some other constant.\cite{Voth:14}
%

%
In MTD simulations the total simulation time required to sample a free energy landscape depends exponentially on the number of CVs.
Larger the volume of the free energy wells, more is the time required to fill them and to see transitions from one well to the other.
Generally, MTD simulations are carried out with 2 or 3 CVs~\cite{mtd:rev:11}. 
%
Although, most of the reactions or structural changes can be sampled using 2 or 3 CVs, one finds several other orthogonal coordinates 
which have hidden barriers, leading to serious errors in the free energies and poor convergence.
Thus inclusion of a few more CVs in sampling would help to accelerate sampling of orthogonal coordinates.
In this spirit,  MTD has been combined with parallel tempering method~\cite{Bussi:06},
and a technique called bias--exchange MTD~\cite{be:mtd:1,be:mtd:2} has been introduced.
%


An ideal technique to compute free energy along a known reaction path is the Umbrella Sampling (US) method,~\cite{us:orig}
where a number of time independent harmonic basing potentials are applied to obtain biased probability distribution of CVs.
Here the umbrella biasing potential placed at $\mathbf s_h$,
\begin{eqnarray}
 \label{eqn:b:us}
W^{\rm b}_h(\mathbf s) = \frac{1}{2} \kappa_h \left ( \mathbf s - \mathbf s_{h} \right )^2, \enspace h=1,\cdots, M 
\end{eqnarray}
is used to construct a set of biased probability distributions $\left \{ \tilde P_h(\mathbf s) \right \}$, 
from $M$ independent MD simulations with varying biases. 
Subsequently, $\left \{ \tilde P_h(\mathbf s) \right \}$ is reweighted to obtain the equilibrium probability distribution $\left \{ P_h(\mathbf s) \right \}$ for all the $M$ windows, and are subsequently combined to get the total distribution $P(\mathbf s) $ through the Weighted Histogram Analysis Method~(WHAM).~\cite{wham:1,wham:2}
The equilibrium free energy surface is then constructed by 
\begin{eqnarray}
F(\mathbf s) = - \frac{1}{\beta} \ln P(\mathbf s) + f
\end{eqnarray}
where $\beta = (k_{\rm B} T)^{-1}$ and $f$ is some arbitrary constant.

\begin{figure}[t]
\begin{center}%
\includegraphics[width=0.8\textwidth]{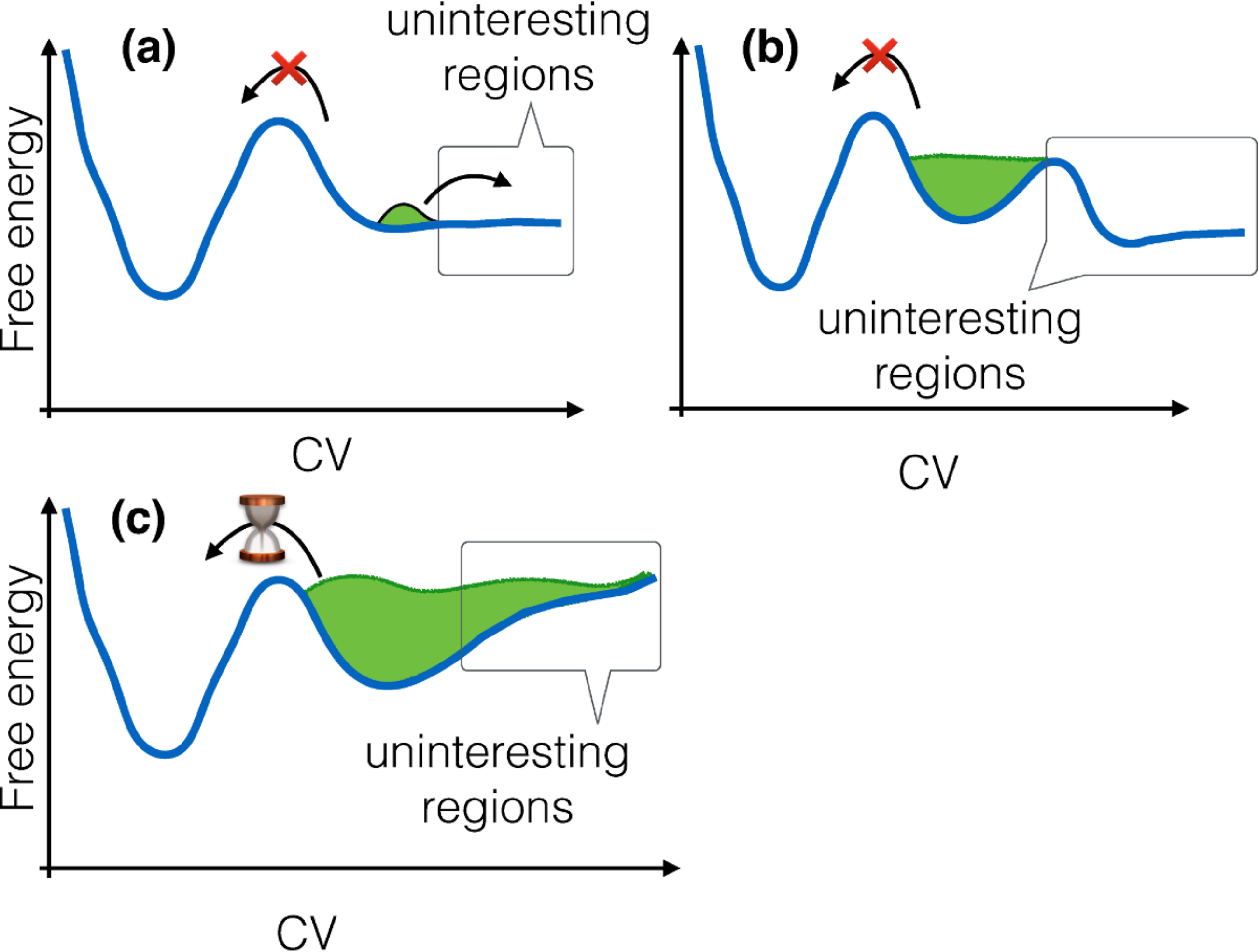}%
\end{center}%
\caption[]{%
\label{fes:cartoon}%
Sketch of free energy surfaces showing three scenarios where traditional MTD simulations fail to
sample efficiently. Green color shows the bias added by MTD. For cases (a) and (b),
transition from reactant well (broad well at the right side) 
to the product well (narrow well at the left) would be failed to observe in a standard MTD. In the case of (c), computational overhead would be very high due to the broad reactant basin.
}
\end{figure}

In \fref{fes:cartoon}, we sketch some practical limitations of standard MTD simulation in sampling flat potentials.
For potentials like those shown in \fref{fes:cartoon}a and b, MTD simulations fail to sample transition from reactant to product wells.
For the case in \fref{fes:cartoon}c, a large number of Gaussian potentials has to be filled in 
 the reactant basin due to its broadness. 
Chemical reactions in systems such as weakly bound Michaelis complexes in enzymes,  weakly bound molecular complexes, 
and  in general A+B type reactions in solutions and gasphase have such free energy topology.
Here, MTD spends most of the computational time in filling uninteresting parts of the free energy wells.
Partly, these problems can be circumvented by using repulsive wall potentials.
%
However, wall potentials introduce boundary effects and require corrections for the boundary effects.~\cite{dePablo:13,Laio:PRB:10} 
%
Often it is difficult to scrutinize whether the minimum observed near the boundary is due to the artifact of the wall potential or actual.
Also, the coordinate that requires controlled sampling using wall potentials has to be defined as CVs (if they are not  
part of the CVs already).
%
Alternatively,  the coordinate along which the free energy is flat or broad can be efficiently sampled using US 
since the sampling range of the coordinate can be controlled 
 in US.
%
%
However, to sample bound orthogonal coordinates with hidden barriers, MTD is ideal because of its  self--guiding nature.
Thus for an efficient sampling of a high dimensional free energy landscape that has features as in \fref{fes:cartoon} along certain CVs, 
a combination of US with MTD, where these methods simultaneously sample orthogonal coordinates, is ideal.

Here we report a procedure to carry out such hybrid simulations and to obtain free energy landscape in the full CV space.
We name this technique of combining US and MTD as Well--Sliced MTD (WS--MTD).
Here $M$ independent MTD  simulations are carried out with the bias
 \begin{eqnarray}
 W^{\rm b}_h (\mathbf s_{\alpha}) + V^{\rm b}_h(\mathbf s_\beta, t), \enspace h=1,\cdots, M
 \end{eqnarray}
to sample the CV space
\[ \mathbf s \equiv \left (\mathbf s_{\alpha},  \mathbf s_{\beta} \right ) \enspace . \]
Here $ W^{\rm b}_h (\mathbf s_{\alpha}) $ and $V^{\rm b}_h(\mathbf s_\beta,t)$ are given by \eref{eqn:b:us} and \eref{eqn:b:mtd}, respectively.
$M$ number of umbrella biases are placed along the $\mathbf s_{\alpha}$ coordinates, and for each of these 
umbrella $h$, we carry out MTD simulation sampling the $\mathbf s_{\beta}$ coordinates.
Each US+MTD simulation samples a slice of the  high--dimensional free energy surface.
 %
Subsequently we combine the biased probability distributions from $M$ different US+MTD simulations to   
construct the total equilibrium distribution and the free energy surface.

Reweighting is not straightforward for MTD, as the biasing potential is time dependent and  the sampling weights
 change with simulation time.
Tiana~\cite{Tiana:08} proposed a strategy to obtain ensemble averages from MTD based on 
a time--dependent reweighting scheme.
%
Employing the convergence property of WT--MTD, a more 
systematic strategy was reported by Bonomi~et al.~\cite{Bonomi:09}
Laio and co--workers have put forward a different reweighting scheme in the framework of bias--exchange MTD.~\cite{be:mtd:2}
Recently, a very simple and efficient reweighting scheme 
 was reported by  
Tiwary and Parrinello~\cite{Tiwary:14} based on time dependent weights directly computable from WT--MTD simulations. 

Combining US with MTD has been reported first by Frilizola and co--workers~\cite{Filizola:12} where
a MTD reweighting proposed by Bonomi et al.~\cite{Bonomi:09}
was used.
US technique has been combined with various sampling techniques; see Ref.~\cite{Kastner:11} for a review and
 Ref.~\cite{US:ITS} for a recent example.
%
%
There were also attempts to do US corrections on the free energy surface obtained from 
MTD.~\cite{ensing05,Sagui:06,Guella:10,Voth:11} 
%

Here we combine US and MTD to sample orthogonal coordinates and reconstruct the free energy surface by a 
combination of Tiwary--Parrinello reweighting scheme and US reweighting within WHAM.
First we carefully study the method on a model potential for which the exact free energy barriers are known.
Then we extend our study to various problems that are part of the ongoing research in our laboratory, where normal
MTD has failed or has poor performance in sampling high dimensional free energy landscape 
due to the problems shown in \fref{fes:cartoon}.
In this respect, we first perform {\em ab initio} MD simulation of formation of cyclobutene from 1,3--butadiene using WS--MTD 
and compare its performance with WT--MTD. 
This serves as an ideal example as its free energy landscape has very broad and deep reactant basin. 
We then applied WS--MTD to model ligand exchange reactions of 
Pd complex in aqueous solution where the standard MTD simulations are known to fail.~\cite{Wacker_ramana}
This example shows how controlled sampling of a high dimensional free energy landscape can be achieved by WS--MTD.
Finally, we demonstrate the efficiency of WS--MTD in modeling an enzymatic reaction using QM/MM method.
Coordination of water molecule to one of the catalytic Zn ions in the active site 
of New Delhi metallo $\beta$--lactamase 1 (NDM--1)~\cite{ravi_nair_NDM} is modeled,
for which our earlier attempts using WT--MTD and non--tempered MTD simulations were
unsuccessful.

\section{Methods and Models}
\label{s:methods}

\subsection{Reweighting Scheme in WS--MTD}
\label{s:methods:reweight}
Here we discuss how to obtain the reweighted distribution from WS--MTD simulations.
%
%
Let us consider a problem where we are interested in computing the free energy surface $F(s_1,s_2)$ by sampling 
the CVs $s_1$ and $s_2$. 
Consider that, the coordinate $s_1$ is sampled using US, while  $s_2$ is simultaneously sampled by a one--dimensional WT--MTD.
If $M$ umbrella potentials are placed along the $s_1$ coordinate, we carry out $M$ independent MTD simulations, each using the Hamiltonian
\begin{eqnarray}
H^{\rm ws}_h(\mathbf R,\mathbf P) =  H^0(\mathbf R,\mathbf P)+ W^{\rm b}_h(s_1) + V^{\rm b}_h(s_2,t)\enspace , \enspace h=1,\cdots, M
\end{eqnarray}
with $W^{\rm b}_h(s_1)$ and $V^{\rm b}_h(s_2,t)$ are given by \eref{eqn:b:us}, and 
\eref{eqn:b:mtd}, respectively, and $H^0$ is the unbiased Hamiltonian for canonical molecular dynamics. 
%
%
For obtaining $F(s_1,s_2)$, we require a strategy to obtain the unbiased probability distribution $P(s_1,s_2)$.
%
%
%

%
When no umbrella bias is present,
under the quasi--stationary limit,
the time dependent probability distribution from a WT--MTD simulation can be written as,
\begin{eqnarray}
\tilde P(\mathbf R,t) = \frac{ \exp \left [ - \beta \left \{ U(\mathbf R)+V^{\rm b} (\mathbf s(\mathbf R),t) \right \} \right ] }{
\int d \mathbf R \, \exp \left [ - \beta \left \{ U(\mathbf R)+V^{\rm b}(\mathbf s(\mathbf R),t) \right \} \right ] }
\end{eqnarray}
and is related to the unbiased probability distribution $P(\mathbf R)$ by~\cite{Bonomi:09}
\begin{eqnarray}
\label{e:p0}
P(\mathbf R) = \tilde P(\mathbf R,t) \exp \left [ \beta \left \{ V^{\rm b}(\mathbf s(\mathbf R),t) - c(t) \right \} \right ] \, 
\end{eqnarray}
where 
\begin{eqnarray}
\label{e:ct}
c(t) = \frac{1}{\beta} \ln \left [ \frac{ \int d \mathbf s \exp [ - \beta F(\mathbf s) ] } 
{ \int d \mathbf s \exp [ - \beta \{ F(\mathbf s)+V^{\rm b}(\mathbf s,t) \} ] } \right ] \enspace .
\end{eqnarray}
%
Evaluation of $c(t)$, however, requires a time--independent free energy  $F(\mathbf s)$, which 
can be computed using the Tiwary--Parrinello time--independent free energy estimator\cite{Tiwary:14} as,
\begin{eqnarray}
\label{e:fs:tiwary}
F(\mathbf s) = -  \alpha V^{\rm b} (\mathbf s, t)  + \frac{1}{\beta} \ln \int d \mathbf s \exp \left [ { \alpha \beta V^{\rm b}(\mathbf s, t)  }  \right ] \enspace ,
\end{eqnarray}
where the time dependency of the first term cancels with that of the second.

In WS--MTD, we construct the time independent probability distribution $P_h^{\rm u}(s_1,s_2)$, by
reweighting the MTD potential, for each umbrella $h$. 
On reaching the quasi--stationary limit, the statistical weight at which $s_2(\mathbf R(t))$ is sampled during MTD (due to MTD bias) 
is 
$\exp[ - \beta \left \{ V_h^{\rm b}(s_2(\mathbf R),t) - c_h(t) \right \} ]$ (see \eref{e:p0}).
Thus,
\begin{eqnarray}
\label{e:p:unb}
P^{\rm u}_h (s_1^\prime,s_2^\prime) =  \frac{ \int_{t_{\rm min}}^{t_{\rm max}} d \tau \, \exp[  \beta \left \{ V_h^{\rm b} (s_2(\tau),\tau) - c_h(\tau) \right \} ] 
                               \enspace \delta(s_1(\tau) - s_1^\prime) \enspace \delta(s_2(\tau) - s_2^\prime)}
                             {\int_{t_{\rm min}}^{t_{\rm max}}  d\tau \, \exp[ \beta \left \{ V_h^{\rm b}(s_2(\tau),\tau) - c_h(\tau) \right \} ]  } \enspace .
\end{eqnarray}
In practice, for each $h$, we compute the above equation by discrete sum over the MTD trajectory of $\{s_1(t),s_2(t)\}$. 
The numerator is computed for bins $(s_1,s_2)$ spanned within a chosen range, while the denominator is independent of the bin value.
We compute integrals for a time series from $t_{\rm min}$ to $t_{\rm max}$ for which the quasi--stationary limit is applicable and 
a proper sampling of $s_2$ is obtained.
%
In WT--MTD, quasi--stationary limit can be thought to be achieved 
when bias is growing slowly and uniformly in the domain of $s_2$ of our interest. 
The bias divergence law~\cite{Branduardi:2012} proves that $c_h(t) \propto \ln(t)$  under this limit  
and thus it is preferred that the reweighting is carried out when this linear relationship is obeyed.~\cite{Tiwary:14}

It may be noted that $P_h^{\rm u} (s_1,s_2)$ is not reweighted for the umbrella bias $W_{h}^{\rm b}(s_1)$.
It is now straightforward to reweight $P_h^{\rm u} (s_1,s_2)$ for the umbrella potential and combine this with WHAM such that
slices of probability densities $P_h^{\rm u}(s_1,s_2), \enspace h=1,\cdots,M$ can be joined to obtain the unbiased distribution 
$P(s_1,s_2)$.
%
WHAM involves minimizing the error in patching the $M$ independent distributions $\{P_{h}^{\rm u}(s_1,s_2) \}$ by 
self--consistently solving
the WHAM equations 
\begin{eqnarray}
\label{e:wham}
P(s_1,s_2) = \frac{\sum_{h=1}^{M}  n_h P_{h}^{\rm u}(s_1,s_2)}
                             {\sum_{h=1}^{M} n_h  \exp[{\beta f_h}] \exp[{ -\beta W_h(s_1)}] } 
\end{eqnarray}
and
\begin{eqnarray}
\label{eqn:fk:u}
 \exp[{-\beta f_h} ] = \int d s_1 \, ds_2 \enspace \exp[ {-\beta W^{\rm b}_h(s_1) }]  P(s_1,s_2) \enspace
\end{eqnarray}
where $n_h$ is the number of configurations sampled in the $h^{\rm th}$ window of the umbrella potential.
The only difference with the usual WHAM equations is that biasing potential along $s_2$ is set to zero.
It is worth noting that \eref{e:wham} assumes that all the windows have nearly the same correlation time.~\cite{wham:1,wham:2}
We have tested this assumption (following the procedure by Hub et al.\cite{Hub:JCTC:10}) for the severe case studied in Section~\ref{s:res:wacker}, and found
that this assumption is a valid one; see Supporting Information for details.

Finally, the free energy surface $F(s_1,s_2)$ is constructed using 
\begin{eqnarray}
\label{e:f}
 F(s_1,s_2) = - \frac{1}{\beta} \ln P(s_1,s_2)
\end{eqnarray}
Thus, we have constructed a two--dimensional free energy surface by $M$ independent 1--dimensional WT--MTD with umbrella restraints.
Although, here we have shown the WS--MTD equations for a two--dimensional case, it is straightforward to generalize this for higher dimensions.
%


WS--MTD has the computational advantage that sampling can be parallelized over the $M$ umbrella windows.
However, the total computational time for WT--MTD increases with $M$ and $t_{\rm max}$ (\eref{e:p:unb}). 
%
%
The total simulation time required for each window, $t_{\rm max}$, depends on the time required to observe $c_h(t) \propto \ln(t)$ behavior and the
time required to sample CVs in MTD after reaching the quasi--stationary limit.
We stress that the benefit of WS--MTD is mainly for the cases where a controlled sampling of a CV is 
required and cannot be achieved in standard MTD.
Especially for the  cases shown in \fref{fes:cartoon}, WS--MTD will be advantageous over WT--MTD.

\subsection{Efficient reweighting of near transition state regions in the MTD CV space}
\label{s:methods:reweight:ts}

To carry out reweighting, $t_{\rm max}$ has to be large enough such that the bias potential is 
changing slowly and uniformly  in the CV--space of our interest.
If one needs to sample two minima separated by a large barrier within the MTD CV space (for a given umbrella), 
the first  trajectory that crosses from the reactant basin to the product basin is insufficient for reweighting the regions near the transition 
state in the CV space, as the sampling near the transition state region is poor and 
has $\dot V^{\rm b}(\mathbf s,t) >> 0$, thus quasi--stationary approximation is not applicable. 
On the other hand, 
for other parts of the CV space where the system has visited several times, bias potential changes 
slowly and uniformly such that reweighting can be carried out.
Ideally, we have to carry out long MTD simulation (for all relevant umbrellas) till the MTD trajectory 
recrosses the two minima multiple times so that bias potential grow slowly near the transition state region. 
However, this is often not practical, especially in {\em ab initio} simulations where the computational overhead for simulating recrossing trajectories is very high,
and for many cases different CVs are required for the reverse reaction. 
Thus, when one is interested only in the forward process (and in computing the forward barrier), 
it is preferred that reweighting is limited to  
the reactant basin and near transition state regions in the CV space, but not the product basin.

We use a simple and straightforward approach to reweight the transition state regions without simulating the recrossing trajectories. 
%
%
If a transition is observed for the first time from reactant well to the product well in the MTD CV space, 
we restart a WT--MTD simulation from an arbitrarily chosen point in the reactant well but using all the bias potential 
$V^{\rm b}(\mathbf s,t)$ accumulated in the previous WT--MTD simulation.
When the reactant to the product transition is observed again, we repeat the same procedure. 
In this way, we increase the sampling near the transition state region, 
and the bias growth rate near the transition state region exponentially decreases towards zero, 
thereby reweighting can be carried out for the transition state region.
Iterations are continued till a satisfactory convergence is achieved for the forward free energy barrier. 
This simple procedure has much less computational overhead and is used in  
the simulations presented here.
In practice, we always start the iterative procedure with the initial structure of the simulation, with velocities
reassigned from Maxwell--Boltzmann distribution.

\subsection{Algorithm}
\label{s:methods:algorithm}

Here we briefly explain the algorithm for the WS--MTD approach. This technique requires no special implementation in a MD code which can
carry out WT--MTD simulation and restrained dynamics simultaneously for different CVs.
The reweighting procedure works as a post processing.

\begin{enumerate}
\item For the chosen range of values of ${\mathbf s_\alpha}$ CVs, place $M$ restraining umbrella potentials. 
For every umbrella, carry out
a WT--MTD simulation sampling the ${\mathbf s_\beta}$ coordinates. 
From these simulations, obtain
the time series of $\mathbf s_\alpha(t),\mathbf s_\beta(t)$, $V^{\rm b}_h(\mathbf s,t)$ for some regular intervals of MTD time $t$.
\item Compute $F_h(\mathbf s)$ using \eref{e:fs:tiwary} for $h=1,\cdots,M$.
\item Compute $c_h(t)$ using \eref{e:ct} for $h=1,\cdots,M$.
\item Plot $c_h(t)$, and based on that choose a time range, $t_{\rm min}$ and $t_{\rm max}$, for which $c_h(t) \propto \ln(t)$, 
and a proper sampling of ${\bf s}_\beta$ has been accomplished.
\item For the time range, construct the MTD--unbiased distributions $P_h^{\rm u}(\mathbf s_\alpha,\mathbf s_\beta), \enspace h=1,\cdots,M$ 
using \eref{e:p:unb}. 
It is crucial that the bin widths are chosen small enough to sample the fluctuations of every umbrella window.
\item Using WHAM based on \eref{e:wham} and \eref{eqn:fk:u}, reweight the umbrella potential as well as combine the $M$ distribution functions to
get $P(\mathbf s_\alpha,\mathbf s_\beta)$. Note that $P_h^{\rm u}(\mathbf s_\alpha,\mathbf s_\beta)$ can be the input for any standard WHAM programs,
but by setting the bias along the $\mathbf s_\beta$ coordinates to zero.
\item Using \eref{e:f}, construct the free energy surface $F(\mathbf s_\alpha,\mathbf s_\beta)$.
\end{enumerate}

\subsection{Computational Details}
\label{s:methods:comput}

\subsubsection{Two Dimensional Model System}
\label{s:methods:model}

For testing the method, we considered a two--dimensional model system which has a broad basin and a narrow basin, 
whose potential is defined as
\begin{eqnarray}
U(x,y) = \sum_{i=1}^{5} U_i^o \exp \left ({-a_i \left [ (x-x_i^o)^2 + (y-y_i^o)^2 \right ]} \right)
\end{eqnarray}
Parameters for the potential are given in \tref{table:uparams}, and the plot of $U(x,y)$ is shown  
in \fref{doublewell_FES}.
The two minima are labeled as {\bf A} (broad minimum) and {\bf B} (narrow minimum), and the barriers for {\bf A}$\rightarrow${\bf B} and {\bf B}$\rightarrow${\bf A} are 9.8~kcal~mol$^{-1}$ and 9.3~kcal~mol$^{-1}$, respectively.

Two different sets of simulations were carried out: (a) WS--MTD simulation with 
$x$ coordinate chosen as the CV ($s_1$) for umbrella sampling 
and the $y$ coordinate as the CV ($s_2$) for one--dimensional WT--MTD sampling;
(b) a two--dimensional WT--MTD simulation with CVs $x$, and $y$ coordinates ($s_1$, and $s_2$,
respectively).
Umbrella potentials were placed along $s_1$ from $-0.05$ to $0.44$~Bohr at intervals of 0.01~Bohr.
Restraining potential $\kappa_h$ for all the umbrella potentials were 
9.9653$\times 10^3$~kcal~mol$^{-1}$~Bohr$^{-2}$.

Mass of the particle was taken as 50~a.m.u. and constant temperature simulation was carried out using Anderson thermostat at 300~K. 
In WS--MTD and two--dimensional WT--MTD simulations, time steps 
for integrating the equations of motion were chosen as 0.12 and 0.17~fs, respectively.
The initial Gaussian height ($w_0$) is 0.63~kcal~mol$^{-1}$ and the hill width parameter $\delta s$ is 
0.005~Bohr.
$\Delta T$ parameter for the WT--MTD simulations was taken as 1200~K.
%

%
%



%
Before starting a WS--MTD, we carried out equilibration for a particular umbrella potential (for 2--5 ps), 
without adding MTD bias.
Initial structure for an umbrella (away from the equilibrium) was chosen from the equilibrated structure of the nearest 
umbrella potential. 
This strategy has been also used for all the other systems studied here.

\subsubsection{1,3--Butadiene to Cyclobutene Reaction}
\label{s:methods:butadiene}

1,3--Butadiene can exist in {\em cis} and {\em trans} form and by an electrocyclic reaction it forms cyclobutene. 
%
%
Free energy surface has a broad reactant basin 
and thus is an ideal problem to demonstrate the efficiency of the WS--MTD method using {\em ab initio} MD.

In order to characterize the {\em cis} and {\em trans} isomers on the free energy landscape and the formation of cyclobutene, we have 
chosen the following CVs (see also \fref{buta_FES}a): a) distance C$_1$--C$_4$, $d[\mathrm C_1-\mathrm C_4]$; b) the difference in the 
distances C$_1$--C$_2$ and C$_2$--C$_3$, $\Delta d[\mathrm C_1-\mathrm C_2-\mathrm C_3]$.
In WS--MTD simulations, $d$[C$_1$--C$_4$] was sampled using US and $\Delta d[\mathrm C_1-\mathrm C_2-\mathrm C_3]$ was sampled using WT--MTD. 
Two--dimensional WT--MTD simulations were also carried out to sample both CVs simultaneously.
 WS--MTD and WT--MTD simulations were performed within the framework of 
{\em ab initio} MD using plane--wave Kohn-Sham density functional theory (DFT) 
employing the CPMD program.~\cite{cpmd1} 
PBE exchange correlation functional~\cite{PBEGGA} with ultrasoft pseudopotential~\cite{Pseudopotential} was used for these calculations. 
A cutoff of 30~Ry was used for the plane--wave expansion of wavefunctions. 
Constant temperature Car--Parrinello~\cite{car85} 
MD at 300~K was carried out using Nos$\mathrm{\grave{e}}$--Hoover chain thermostats.~\cite{nhc}
A time step of 0.096~fs was used to integrate the equations of motion. 
A mass of 600~a.u. was assigned to the orbital degrees of freedom.
System was taken in a cubic supercell of side length 15~{\AA}.
Extended Lagrangian MTD approach was used here with a harmonic coupling constant of 1.0~a.u. 
to restrain the CVs and collective coordinates, and a mass of 50.0~a.m.u. was assigned to all the CVs.
Langevin thermostat with a friction coefficient of 0.001~a.u. was used to maintain the CV temperature to 300~K.
In WT--MTD simulations, $w_0=0.6$~kcal~mol$^{-1}$ and $\delta s=$0.05~a.u. were taken. 
MTD bias was updated every 19~fs. 
Two independent two--dimensional WT--MTD simulations were performed using $\Delta T=3000$~K and 25000~K.
In US, windows were placed from 1.5~\textrm{\AA} to 3.9~\textrm{\AA} at an interval of 0.05~\textrm{\AA} with $\kappa_{\rm h} = 1.57 \times 10^3$~kcal~mol$^{-1}$~\textrm{\AA}$^{-2}$.
In the case of WS--MTD, $\Delta T=3000$~K was used while all the other MTD parameters were the same as in the case of the normal WT--MTD. 

\subsubsection{Controlled Sampling of Ligand Exchange in a Pd--Allyl Alcohol Complex in Aqueous Solution}
\label{s:methods:wacker}

%
%
Here we are interested to compute the free energy barrier for the reaction {\bf WA1}$\rightarrow${\bf WA2} (\fref{wacker_FES}a) 
in aqueous solution. 
The purpose of this simulation is to achieve a controlled sampling, in particular, to avoid sampling the rapid ligand exchanges of {\em trans} Cl 
with solvent molecules to form {\bf WA3}.~\cite{Wacker_ramana} 
To simulate this process we have chosen four CVs: a) coordination number Pd to all the oxygen atoms of water molecules, $C\mathrm{[Pd-O_w]}$; b) 
coordination number of Pd to Cl atom that is {\em trans} to the coordinated allyl alcohol, $C\mathrm{[Pd-Cl_{trans}]}$; 
c) coordination number of Pd to the two C atoms of allyl alcohol to which it is coordinated, $C\mathrm{[Pd-C]}$; 
d) coordination number of Pd to the Cl$^-$ in  solution, $C\mathrm{[Pd-Cl]}$.
{\em Ab initio} MD simulations were performed with the same technical details as mentioned previously.
A periodic cubic box having the side length of 12.0~{\AA} containing the {\bf WA1} complex, one Cl$^-$, 
and 53 water molecules were taken.
The equilibrated structure of this system was taken from our previous study in Ref.~\cite{Wacker_ramana}.

We carried out WS--MTD simulations by sampling the CVs $C\mathrm{[Pd-O_w]}$ and $C\mathrm{[Pd-Cl_{trans}]}$ by US, and
the CVs $C\mathrm{[Pd-C]}$ and $C\mathrm{[Pd-Cl]}$ by WT--MTD.
In these simulations, the umbrella windows along the CV $C\mathrm{[Pd-Cl_{{trans}}]}$ were placed at 0.86 and 0.74,
with $\kappa_h$ 251~kcal~mol$^{-1}$ and 502~kcal~mol$^{-1}$, respectively. 
The position of the 
umbrella window along the CV $C\mathrm{[Pd-O_{w}]}$ was fixed at 1.24.
The $\kappa_h$ for the umbrellas along $C\mathrm{[Pd-O_w]}$ was 21.3~kcal~mol$^{-1}$.  
The position and $\kappa_h$ for the umbrella restraints were based on the distribution of these CVs at equilibrium in
the {\bf WA1} state.

We carried out the extended Lagrangian MTD where a harmonic coupling constant of 2.0~a.u. was used to 
restrain CVs and collective coordinates,  and a CV mass
of 50.0~a.m.u was assigned. 
Temperature of the CV was maintained  to 300$\pm$ 200~K by a direct velocity scaling. 
%
%
MTD biasing potential was updated every 29~fs and the MTD parameters $w_0=0.62$~kcal~mol$^{-1}$, 
and $\delta s= 0.05$ were taken.  
$\Delta T=3000$~K was chosen for these simulations.

As we are only interested in the forward barrier along the MTD CVs, we have used the 
 strategy presented in Section~\ref{s:methods:reweight:ts}.
%
%
Moreover, to trace the minimum energy pathway on the reconstructed five--dimensional free energy surface,
we used the string method.\cite{string_method} 

\subsubsection{Free Energy for Water Coordination to the Active Site of NDM--1 by QM/MM Simulations}
\label{s:methods:ndm1}

\begin{figure}[t]
\begin{center}%
\includegraphics[width=0.7\textwidth]{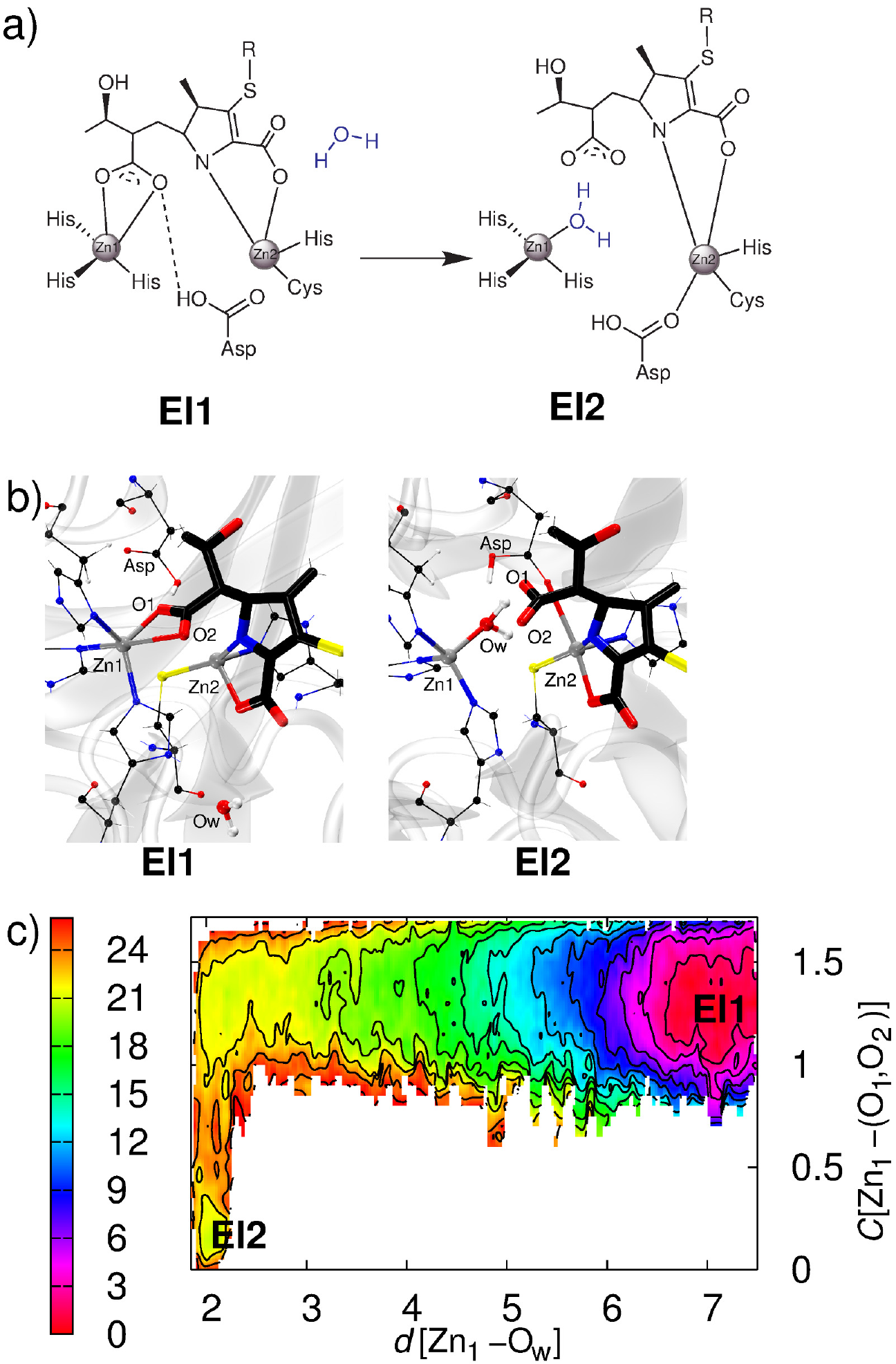}%
\end{center}%
\caption[]{%
\label{ndm_FES}%
(a) Structures of ({\bf EI1}) and ({\bf EI2}) with active site water molecule shown in blue color.
(b) Snapshots of {\bf EI1}, and {\bf EI2}; color code: Zn (gray), C (black), O (red), N (blue), S
(yellow), and H (white). Active site residues and the attacking water molecule are in CPK representation,
while the drug molecule is shown in stick representation.
(c)Reconstructed free energy computed from WS--MTD.
Free energy values are in kcal~mol$^{-1}$ relative to the free energy of the minimum ({\bf EI1}).
Contour values are drawn from 1 to 26 kcal~mol$^{-1}$ for every 2~kcal~mol$^{-1}$.
$d$[Zn$_1$--O$\rm_w$] is in \textrm{\AA}.

}
\end{figure}

%
%
To sample the reaction from {\bf EI1}$\rightarrow${\bf EI2} two CVs were chosen (\fref{ndm_FES}b): 
a) distance between Zn$_1$ and to the nearest water molecule (as identified in our earlier work~\cite{ravi_nair_NDM}), $d\mathrm{[Zn_1-O_w]}$;
b) coordination number of Zn$_1$ to the carboxylate oxygen atoms O$_1$ and O$_2$, $C\mathrm{[Zn_1-(O_1,O_2)]}$. 
The first coordinate was chosen as the US CV as we are expecting a broad and deep free energy basin along this coordinate, 
while the other coordinate was chosen for the WS--MTD sampling. 
Umbrella potentials were placed from 7.4~\textrm{\AA} to 1.9~\textrm{\AA} at an interval of 0.1~\textrm{\AA} 
with $\kappa_{\rm h} = 4.5 \times 10^2$~kcal~mol$^{-1}$~\textrm{\AA}$^{-2}$.

Hybrid QM/MM simulations were performed using the CPMD/GROMOS interface~\cite{laio02a} 
as implemented in the CPMD package.
The equilibrated structure of {\bf EI1} was taken from our previous work~\cite{ravi_nair_NDM}; see  (\fref{ndm_FES}).
Two Zn ions, meropenem drug, side chains of His$_{120}$, His$_{122}$, Asp$_{124}$, His$_{189}$, Cys$_{208}$, and His$_{250}$ 
 and the water molecule near the active site were treated quantum mechanically (QM).
Rest of the protein  and the solvent molecules were treated by molecular mechanics (MM), and were
taken in a periodic box of size 69.0$\times$63.2$\times$66.8~\textrm{\AA}$^{3}$. 
The whole protein, including the QM part, and the solvent molecules were free to move during the MD simulations.
The QM part was treated using the plane--wave DFT with Ultrasoft pseudopotentials and with a plane--wave cutoff of 30~Ry.
A QM box size of 27.5$\times$27.8$\times$22.8~\textrm{\AA}$^{3}$ was chosen. 
Whenever a chemical bond has to be cleaved to define partition between QM and MM parts, boundary atoms are capped using hydrogen atoms.
Capping hydrogen atoms were introduced between C$_\beta$ and C$_\gamma$ atoms of His$_{120}$, His$_{122}$, His$_{129}$, and His$_{250}$
and C$_\alpha$ and C$_\beta$ atoms of Asp$_{124}$, and Cys$_{208}$.

Car--Parrinello MD was carried out for the QM part.
Time step for the integration of the equations of motion was 0.145~fs. 
A mass of 700 a.u. was assigned to the orbital degrees of freedom and Nos{\'e}--Hoover chain thermostats~\cite{nhc} 
 were used to perform $NVT$ ensemble simulations at 300~K.

Extended Lagrangian variant of MTD was used where harmonic coupling constant was taken as 2.0~a.u. and the CV mass was set to 50.0~a.m.u.
CV temperature was maintained to 300~K by coupling to Langevin thermostat with a frictional coefficient of 0.001~a.u.
Gaussian potentials were updated every 29~fs and the MTD parameters $w_0=0.62$~kcal~mol$^{-1}$, $\delta s= 0.05$ and $\Delta T= 7500$~K were taken.
We followed the strategy described in Section~\ref{s:methods:reweight:ts} for reweighting the CV space near the transition state along the MTD CV.

\section{Results and Discussion}
\label{s:res}

First we benchmark the accuracy of WS--MTD method by sampling a two--dimensional model system 
where the free energy barriers are exactly known (Section~\ref{s:res:model}).
To demonstrate the efficiency of WS--MTD simulation over the WT--MTD simulation in sampling broad and deep free energy wells,
we model the cyclization reaction of 1,3--butadiene using {\em ab initio} MD (Section~\ref{s:res:butadiene}).
In Section~\ref{s:res:wacker}, we demonstrate a controlled sampling of a high--dimensional surface 
 using WS--MTD, which is otherwise difficult using normal MTD.
Subsequently, we present an example where WS--MTD is used together with QM/MM methods to sample 
the broad free energy surface of an 
enzymatic reaction, for which normal MTD simulations have failed.
%
%

\subsection{Two Dimensional Model System}
\label{s:res:model}

\begin{figure}[t]
\begin{center}%
\includegraphics[width=0.8\textwidth]{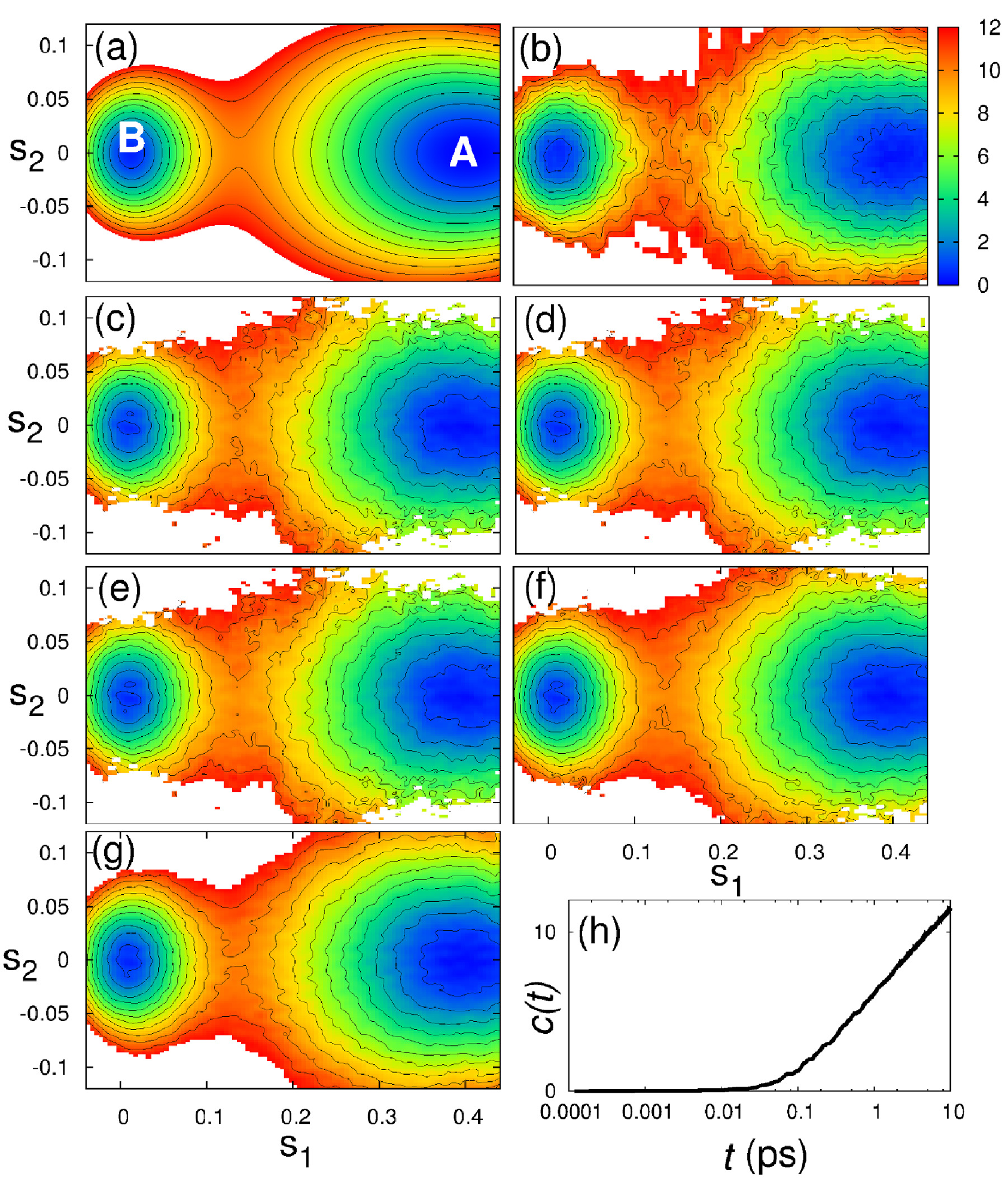}%
\end{center}%
\caption[]{%
\label{doublewell_FES}%
Two--dimensional double well potential: (a) Potential energy surface; (b) free energy surface from two--dimensional WT--MTD;  Free energy
surfaces from WS--MTD simulations for ($t_{\rm min},t_{\rm max}$) values (in ps): 
(c) (0.1,1.4), (d) (0.1,2.8), (e) (0.2,1.4), (f) (0.2,2.8), (g) (1.4,11.4); (h) $c(t)$ plot for one of the windows in
the WS--MTD simulation where $t$ axis is in log--scale. Free energy values are in kcal~mol$^{-1}$ relative to the free energy of 
the broad minimum ({\bf A}). $c(t)$ is in kcal~mol$^{-1}$. Contour values are drawn from 1.0 to 12.0 kcal~mol$^{-1}$ at 1~kcal~mol$^{-1}$ intervals.
}
\end{figure}

We carried out a two--dimensional WT--MTD simulation starting from the minimum {\bf A} using the CVs $s_1(\equiv x)$ and $s_2(\equiv y)$.
Several re-crossings were observed in the WT--MTD simulation, 
while a proper convergence in the barriers was observed by the third recrossing at 86.7~ps. 
The converged free energy barriers for {\bf A}$\rightarrow${\bf B} and {\bf B}$\rightarrow${\bf A}
are 9.8 and 9.3~kcal~mol$^{-1}$, respectively, and are identical to the exact barriers from the potential energy surface (\fref{doublewell_FES}a).
Free energy surface from this simulation is shown in \fref{doublewell_FES}b.
We then carried out WS--MTD simulations, by sampling $s_1$ and $s_2$ using  
US and one--dimensional WT--MTD simulation, respectively.
During these simulations, $c(t)$ values for all the 50 umbrella windows were computed according to \eref{e:ct}.
For all the umbrellas,  $c(t)$--$t$ plots were nearly identical, and one of the plots is shown
in \fref{doublewell_FES}h. 
$c(t) \propto \ln(t)$ behavior was observed for $t> 0.1$~ps for all the umbrella windows.
Thus, for reweighting the MTD simulations using \eref{e:p:unb},
we tested various values of $t_{\rm min}$ and $t_{\rm max}$ greater than 0.1~ps, and the resulting free energy surfaces were analyzed
for convergence in free energy barriers.
For $t_{\rm min}=0.1$~ps and $t_{\rm max}=0.6$~ps, the reconstructed free energy surface was quite noisy, and the free energy barriers are having an error 
upto 0.4~kcal~mol$^{-1}$ (\tref{t:mtd:conv}).
On the other hand, for $t_{\rm max}=1.4$~ps, free energy barriers for {\bf A}$\rightarrow${\bf B} and {\bf B}$\rightarrow${\bf A}
were 9.8 and 9.2~kcal~mol$^{-1}$, respectively, which is in good agreement with the reference free energy barriers; see also \fref{doublewell_FES}c.
The topology of the relevant parts of the free energy surface 
is also well reproduced.
For higher $t_{\rm max}$ values, the free energy barriers converge to the exact values (\tref{t:mtd:conv}). 
The same convergence behavior was also observed for $t_{\rm min}>0.1$~ps.
It is also interesting to note that $t_{\rm min}=0.0$ and $t_{\rm max}=1.4$ also gave reasonably accurate free energy barriers; 
see also the figure in the
supporting information where error in the free energy barriers as a function of cumulative simulation time is shown for both
WS--MTD and WT--MTD.
Thus, performances of WS--MTD and WT--MTD are nearly the same, but
the advantage of WS--MTD will become more obvious in the realistic examples discussed later.
%
In cases with large $t_{\rm max}$, parts of the surfaces with higher free energy are better explored; see \fref{doublewell_FES}d,f,g.
However, higher free energy regions are not often interesting and  $t_{\rm max}$ could be chosen based on convergence in the 
computed free energy barriers in more complex realistic systems.

\subsection {1,3--Butadiene to Cyclobutene Reaction}
\label{s:res:butadiene}

\begin{figure}[t]
\begin{center}%
\includegraphics[width=0.85\textwidth]{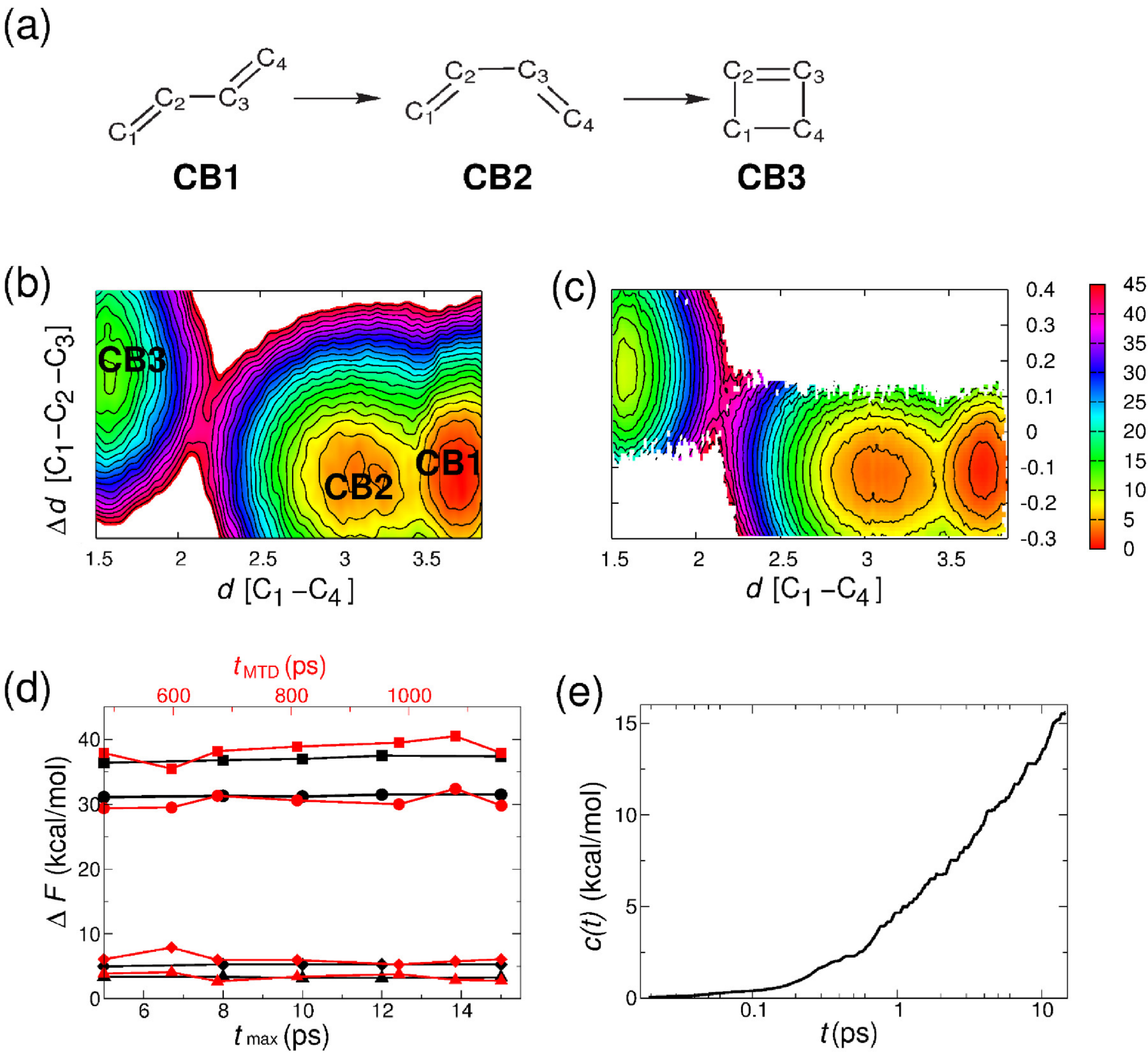}%
\end{center}%
\caption[]{%
\label{buta_FES}%
(a) Structure of {\em trans}-1,3-butadiene ({\bf CB1}), 
{\em cis}-1,3-butadiene ({\bf CB2}), 
and cyclobutene ({\bf CB3}). 
Reconstructed free energy surface for 
1,3-butadiene to cyclobutene reaction
computed from (b) two--dimensional WT--MTD ($t_{\rm MTD}$=1.2~ns), and 
(c) WS--MTD ($t_{\rm min}$=0.0, $t_{\rm max}$=15~ps). 
Free energy values are in kcal~mol$^{-1}$ relative to the free energy of the minimum 
({\bf CB1}). 
Contour values are shown from 1.0 to 45.0 kcal~mol$^{-1}$ for every 2~kcal~mol$^{-1}$. CVs are in \textrm{\AA}. 
%
(d) Various free energy barriers (color code: WS--MTD (black), WT--MTD (red)) on the free energy landscape computed for different $t_{\rm max}$ values from WS--MTD simulation (bottom axis)
together with the corresponding free energy barriers from two--dimensional WT--MTD with increasing MTD time, 
$t_{\rm MTD}$ (top axis). Here symbols $\blacksquare$, $\bullet$, $\blacklozenge$, $\blacktriangle$ represent free energy barriers for 
{\bf CB2}$\rightarrow${\bf CB3}, {\bf CB3}$\rightarrow${\bf CB2}, {\bf CB1}$\rightarrow${\bf CB2}, and {\bf CB2}$\rightarrow${\bf CB1} respectively.
%
(e) $c(t)-t$ plot for one of the windows in the WS--MTD simulation, where the $t$ axis is in log--scale.
}
\end{figure}

Here we carried out a WS--MTD simulation starting with a 1,3--butadiene structure (\fref{buta_FES}a).
The converged reconstructed free energy surface from this simulation is shown in (\fref{buta_FES}c), where three minima 
{\bf CB1}, {\bf CB2}, {\bf CB3}  
could be observed  corresponding to {\em trans}-buta-1,3-diene, {\em cis}-buta-1,3-diene and cyclobutene, respectively.
In WS--MTD, the converged barriers for going from 
{\bf CB1}$\rightarrow${\bf CB2} and {\bf CB2}$\rightarrow${\bf CB1}  
are 5.3 and 3.3~kcal~mol$^{-1}$, respectively.
The free energy barriers for {\bf CB2}$\rightarrow${\bf CB3} and {\bf CB3}$\rightarrow${\bf CB2}
are 37.4 and 31.4~kcal~mol$^{-1}$, respectively (\fref{buta_FES}c,d). 
Convergence in all the free energy barriers is achieved with $t_{\rm min}=0.0$~ps and $t_{\rm max}=5.0$~ps  
(for all the umbrella potentials); see Supporting Information for a more detailed analysis.

To compare the performance of a two--dimensional WT--MTD sampling of both CVs, 
we carried out two independent WT--MTD simulations with two different $\Delta T$ parameters.
In the first simulation, $\Delta T$ was taken as 3000~K, which is the same as that was used for WS--MTD.
In this case, even after 1.2~ns of the simulation, no crossing for {\bf CB2}$\rightarrow${\bf CB3} was seen (see Supporting Information).
%
%
The failure of WT--MTD is due to the presence of broad and deep reactant well and small $\Delta T$ parameter. 
%
Thus a second WT--MTD simulation was then carried out with $\Delta T= 25000$~K, where we were successful in seeing  several re-crossings 
 from {\bf CB2}$\rightarrow${\bf CB3} within 1.2~ns.  
Although free energy barriers for 
{\bf CB1}$\rightarrow${\bf CB2} and {\bf CB2}$\rightarrow${\bf CB1}  
have converged systematically (due to the small barriers separating them), 
 the free energy barriers for 
 {\bf CB2}$\rightarrow${\bf CB3} and {\bf CB3}$\rightarrow${\bf CB2} 
are not converged to the extent in WS--MTD simulations even after 1.2~ns (\fref{buta_FES}d). 
%
%
%
%
%
%
Free energy barriers computed from WS--MTD are well converged within 5.0~ps simulation, but not for WT--MTD even after 1.2~ns.
Since we used 49 windows and $t_{\rm max}=5$~ps per umbrella, the total computational time invested for a well converged result in WT--MTD is 245~ps. 
Also, note that all the 49 windows were running in parallel in our computations and the clock time to complete a converged WS--MTD was equivalent to the
clock time required  required for 
simulating 5.0~ps trajectory of a two--dimensional WT--MTD on same number of processors allocated per umbrella.
%
%

We thus clearly show for a realistic system, that WS--MTD has better performance over the normal WT--MTD 
in sampling free energy surfaces with broad and deep basins.
%

\subsection {Controlled Sampling of Ligand Exchange in a Pd--Allyl Alcohol Complex in Aqueous Solution}
\label{s:res:wacker}

\begin{figure}[t]
\begin{center}%
\includegraphics[width=0.85\textwidth]{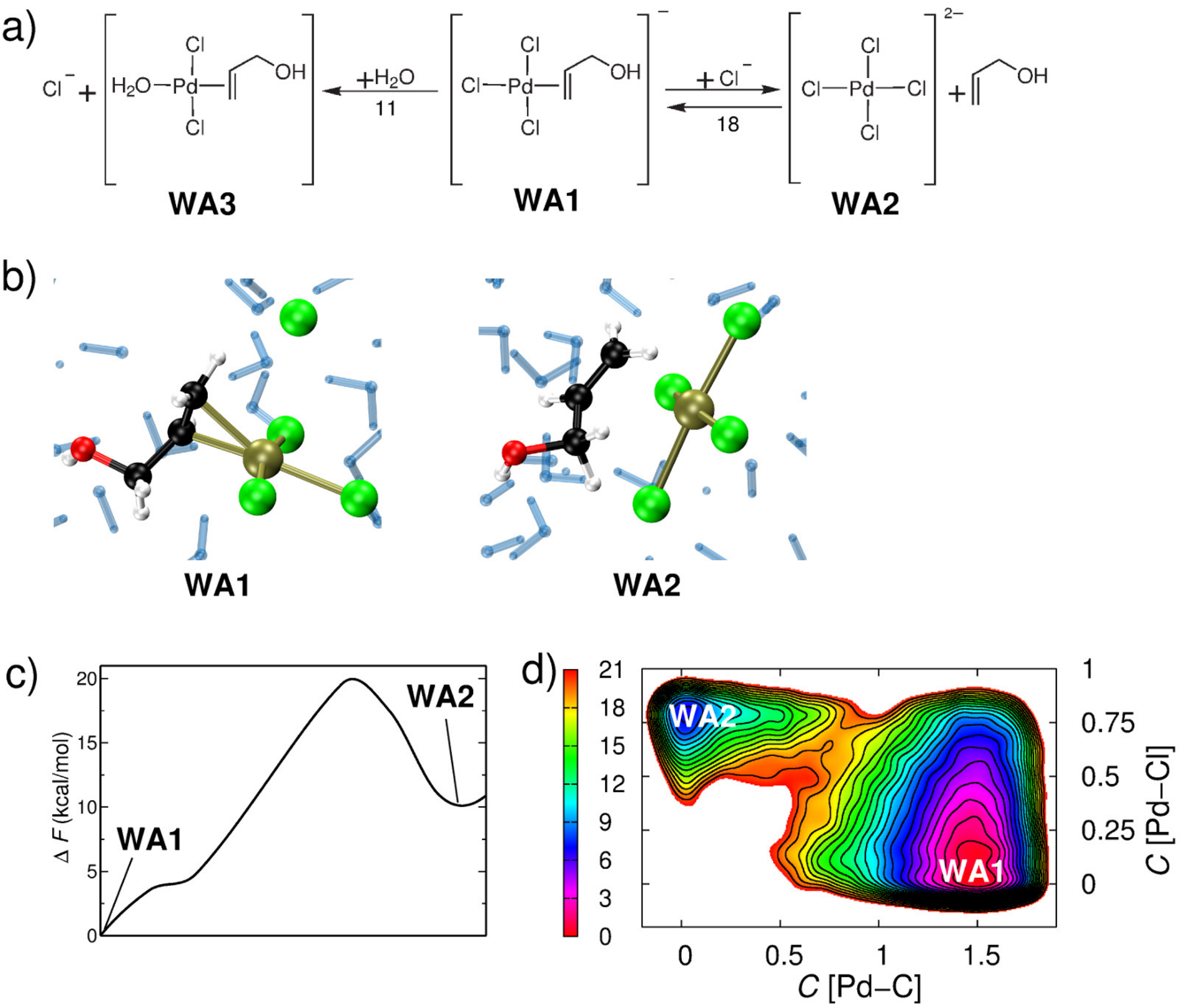}%
\end{center}%
\caption[]{%
\label{wacker_FES}%
(a) Structure of Pd complex with allyl alcohol ({\bf WA1}), PdCl$_4^{2-}$ ({\bf WA2}) and Pd complex with water ({\bf WA3}). 
Free energy barriers in kcal~mol$^{-1}$ are indicated over arrows, as computed 
from our previous study.\cite{Wacker_ramana} 
(b)Snapshots from the simulation showing the complex of Pd with allyl alcohol ({\bf WA1}), and PdCl$_4^{2-}$ ({\bf WA2}); 
color code: Pd (tan), Cl (green), C (black), O(red), H(white) and solvent water molecules in blue stick representation.
(c) Minimum energy pathway traced on the converged five--dimensional free energy landscape. 
(d) The non reweighted reconstructed free energy surface 
{\bf WA1}$\rightarrow${\bf WA2} 
visualized along the MTD CVs for
$C\mathrm{[Pd-Cl_{trans}]}=0.74$ 
and 
$C\mathrm{[Pd-O_{w}]}=1.24$. 
Free Energy values are in kcal~mol$^{-1}$ relative to the free energy of the minimum 
({\bf WA1}).  
Contour values are drawn from 1.0 to 20.0 kcal~mol$^{-1}$ for every 1~kcal~mol$^{-1}$. 
%
%
}
\end{figure}

In our earlier work on the investigation of the Wacker oxidation of allyl alcohol~\cite{Wacker_ramana} 
in aqueous solution
we were interested in modeling 
{\bf WA1}$\rightarrow${\bf WA2}. 
However, we were unable to simulate this step using normal MTD
due to rapid ligand exchange reaction of the {\em trans}--Cl  (Cl$_{\rm trans}$) with water molecules from
solution, as a result of the strong {\em trans}--directing nature of the olefinic group (\fref{wacker_FES}a).
Different wall potentials at different values of Pd--Cl$_{\rm trans}$ coordination number were resulting in
different free energy barriers and thus a reliable estimation could not be made.
Thus, our earlier work\cite{Wacker_ramana} has used the equilibrium constant for 
{\bf WA2}$\leftrightharpoons${\bf WA1}  
from experiment and the barrier for  {\bf WA2}$\rightarrow${\bf WA1} 
computed from MTD
to estimate the {\bf WA1}$\rightarrow${\bf WA2} 
barrier as $\approx$19~kcal~mol$^{-1}$.

In order to compute the free energy barrier for {\bf WA1}$\rightarrow${\bf WA2}, we have sampled the
$C\mathrm{[Pd-C]}$, $C\mathrm{[Pd-Cl]}$,
$C\mathrm{[Pd-Cl_{trans}]}$ and $C\mathrm{[Pd-O_w]}$.
The controlled sampling of the $C\mathrm{[Pd-Cl_{trans}]}$ and $C\mathrm{[Pd-O_w]}$ CVs was achieved by using US
while the other CVs were sampled by a two--dimensional WT--MTD.
The restraining values of US coordinates were determined form equilibrium simulation, and the restraining potential
was computed from the standard deviation of the probability distribution of these CVs without any MTD bias.
Along the $C\mathrm{[Pd-O_w]}$ coordinate, we do not want to sample other umbrella windows in order to prevent water coordination and thus
the reaction to 
{\bf WA3}. 
Based on our chemical intuition and from our previous experience,\cite{Wacker_ramana}
we interpret that $C\mathrm{[Pd-Cl_{trans}]}$ could vary  from the equilibrium value along the reaction, 
because when external Cl$^-$ approaches axially and coordinate to Pd,
an increase in the Pd--Cl$_{\rm trans}$ distance is expected.
Thus we sampled another overlapping window
for the $C\mathrm{[Pd-Cl_{trans}]}$ CV  at a lower value
(Section~\ref{s:methods:wacker}). 
In this way, we carried out WS--MTD simulation to sample only the crucial part of the five--dimensional free energy landscape. 
Minimum energy pathway on this high--dimensional landscape was then obtained by
the string method (\fref{wacker_FES}c).~\cite{string_method} 

Since reactant and product basins are present in the MTD CV space
(see also \fref{wacker_FES}d)
and that we are only interested in obtaining the forward barrier 
i.e. for the reaction {\bf WA1}$\rightarrow${\bf WA2},
we used the iterative approach for reweighting as discussed in 
Section~\ref{s:methods:reweight:ts}.
The convergence in free energy barriers with number of iterations is given in \tref{table:wacker_walk}. 
A reasonable convergence was obtained for the third iteration, and free energy barrier for 
{\bf WA1}$\rightarrow${\bf WA2} 
is computed as 20~kcal~mol$^{-1}$ (\fref{wacker_FES}c).
This is in excellent agreement with the estimated barrier of 19~kcal~mol$^{-1}$ in our earlier work.~\cite{Wacker_ramana}

%
By presenting this non--trivial example, we demonstrate that WS--MTD can be applied for a controlled sampling of a  
 high--dimensional free energy landscape of a complex reaction.
Moreover, the iterative scheme for reweighting the near transition state regions 
is demonstrated for a realistic system.

\subsection {Free Energy for Water Coordination to the Active Site of NDM--1 by QM/MM Simulations}
\label{s:res:ndm1}

In this section, we study a problem which is part of our ongoing research in elucidating 
the mechanistic details of antibiotic resistance in NDM--1.~\cite{ravi_nair_NDM}
In the intermediate stage of one of the proposed mechanisms for the hydrolysis of the acyl--enzyme complex within the active site 
requires a water molecule to enter the active site and coordinate to the Zn$_1$ site, further leading
to the dissociation of drug--carboxylate group coordination to Zn$_1$ and subsequent protonation of $\beta$--lactam N (see \fref{ndm_FES}a).~\cite{ravi_nair_NDM}
We attempted nearly 10 different MTD simulations with various collective variables 
to model the entry of a QM water to the active site and its coordination to Zn$_1$, however,
they all failed to simulate the reaction, and hinted a broad reactant basin.
%
%
%

Thus, we addressed this problem using WS--MTD by US along $d\mathrm{[Zn_1-O_w]}$ and WT--MTD sampling
along $C\mathrm{[Zn_1-(O_1,O_2)]}$.
The US windows at $d\mathrm{[Zn_1-O_w]}> 2.5$~{\AA} 
was only simulated till  
the $\dot V^{\rm b}(s,t) \approx 0$ limit
is achieved in the relevant values of the $C\mathrm{[Zn_1-(O_1,O_2)]}$ coordinate.
MTD biases are added for the windows with $d\mathrm{[Zn_1-O_w]}< 2.5$~{\AA} till a transition is observed along  $C\mathrm{[Zn_1-(O_1,O_2)]}$ 
or till the total bias is equal to 25~kcal~mol$^{-1}$ along the $C\mathrm{[Zn_1-(O_1,O_2)]}$ coordinate.
If a transition is observed along the $C\mathrm{[Zn_1-(O_1,O_2)]}$ coordinate, we carried out the iterative approach for reweighting the
transition state regions of the CV space as mentioned in 
Section~\ref{s:methods:reweight:ts}. 

The reconstructed free energy surface thus obtained 
is given in \fref{ndm_FES}c.
The converged free energy barrier for this reaction is 23.6~kcal~mol$^{-1}$.
From the topology of the surface, it is now clear why the normal MTD simulations failed:
the
free energy basin is deep and flat along the $d\mathrm{[Zn_1-O_w]}$ coordinate, and the minimum of 
the reactant basin is at $d\mathrm{[Zn_1-O_w]}\approx 7$~{\AA}.
Free energy surface also indicates that water coordination to Zn$_1$ and  dissociation of Zn$_1$--carboxylate bond occur in tandem.

\section{Conclusions}
Based on a combination of Tiwary--Parrinello and US reweighting techniques within WHAM, we propose 
the WS--MTD strategy for reconstructing complex free energy landscapes by sampling them by slices.
Having the advantage of controlling the sampling along a few selected coordinates, WS--MTD 
aids in efficient sampling of flat, broad and unbound free energy wells with hidden orthogonal barriers.
%
%
We also propose a simple strategy to reweight MTD trajectories 
for the regions near the transition state along the MTD CVs
in order to obtain converged forward 
barriers without the need of sampling recrossing trajectories.
The WS--MTD also has the advantage of parallelization over umbrellas.
This feature enables sampling of high--dimensional landscapes in affordable clock time
employing more computational resources in parallel.
WS--MTD does not require any special implementation in MD codes that 
can simultaneously perform restrained dynamics and WT--MTD.

We carried out careful study on the accuracy of the method by sampling a two dimensional 
model potential where the free energy barriers are exactly known.
Free energy barriers were found to converge systematically to the exact values with increasing
$t_{\rm max}$.
Several applications using the WS--MTD  method were then presented.
Efficiency of WS--MTD in sampling broad and flat free energy surface is demonstrated
by studying the cyclization of 1,3--butadiene using {\em ab initio} MD. 
WS--MTD yields converged free energy barriers with a much less computational
cost (by a factor of 5 at least for this problem) than a normal MTD.
By sampling the five--dimensional free energy surface for a ligand exchange reaction in a 
Pd--allyl alcohol complex solvated in water, we demonstrate 
the application of WS--MTD in controlled sampling of a high dimensional surface.
The modeled ligand exchange reaction was earlier reported to 
fail using normal MTD.
Also we showed the application of an iterative approach to reweight the
MTD trajectories near the transition state regions on the CV space 
without the need of simulating multiple recrossing trajectories.
Finally, we apply WS--MTD to sample an enzymatic reaction using QM/MM technique, where the
coordination of a catalytic water to the active site of NDM--1 enzyme is simulated.
For this problem, WS--MTD could efficiently sample broad and deep free energy basins, which was otherwise
unable to achieve in a normal MTD after several attempts.

Since the method is practically applicable in {\em ab initio} MD and QM/MM MD simulations, we believe that
this would be useful for sampling high-dimensional free energy landscape of complex chemical reactions.
The technique would be beneficial for various problems in chemistry and biology such as
A+B type reactions in weakly bound molecular complexes, reactions in Michaelis complexes, and substrate binding in enzymes to name a few,
where one encounters broad and deep free energy basins.
%

%

\section*{Acknowledgments}
Authors gratefully acknowledge the discussions with Dr. Pratyush Tiwary, Mr. Chandan Kumar Das, Dr. Venkataramana Imandi,  and Dr. Ravi Tripathi.
Authors also thank IIT Kanpur for availing the computing resources. 
NN acknowledges IIT Kanpur for the support through P.~K. Kelkar Young Faculty Research Fellowship. 
SA thanks UGC for Ph.~D fellowship.

\section*{Supporting Information} 
Supporting information has the following data: a) table analyzing the free energy convergence in WS--MTD and WT--MTD simulations of 1,3--butadiene to cyclobutene reaction; 
b) a plot showing the convergence of error as a function of cumulative time for WS--MTD and WT--MTD for the two--dimensional model potential; c) details on the computation of
correlation time for the two umbrella windows used for the problem in Section 3.3. 


\begin{thebibliography}{10}

\bibitem{Laio:PNAS:02}
A.~Laio, M.~Parrinello, \emph{Proc. Natl. Acad. Sci.} \textbf{2002}, \emph{99},
  12562--12566.

\bibitem{Iannuzzi:03}
M.~Iannuzzi, A.~Laio, M.~Parrinello, \emph{Phys. Rev. Lett.} \textbf{2003},
  \emph{90}, 238302.

\bibitem{mtd:rev:11}
A.~Barducci, M.~Bonomi, M.~Parrinello, \emph{WIREs Comput.~Mol.~Sci.}
  \textbf{2011}, \emph{1}, 826--843.

\bibitem{Luigi:12}
L.~Sutto, S.~Marsili, F.~L. Gervasio, \emph{WIREs: Comput. Mol. Sci.}
  \textbf{2012}, \emph{2}, 771--779.

\bibitem{Gervasio:08}
A.~Laio, F.~L. Gervasio, \emph{Rep.~Prog.~Phys.} \textbf{2008}, \emph{71},
  126601.

\bibitem{mtd:well:08}
A.~Barducci, G.~Bussi, M.~Parrinello, \emph{Phys. Rev. Lett.} \textbf{2008},
  \emph{100}, 020603.

\bibitem{Voth:14}
J.~F. Dama, M.~Parrinello, G.~A. Voth, \emph{Phys.~Rev.~Lett.} \textbf{2014},
  \emph{112}, 240602.

\bibitem{Bussi:06}
G.~Bussi, F.~L. Gervasio, A.~Laio, M.~Parrinello, \emph{J. Am. Chem. Soc.}
  \textbf{2006}, \emph{128}, 13435--13441.

\bibitem{be:mtd:1}
S.~Piana, A.~Laio, \emph{J. Phys. Chem. B} \textbf{2007}, \emph{111},
  4553--4559.

\bibitem{be:mtd:2}
F.~Marinelli, F.~Pietrucci, A.~Laio, S.~Piana, \emph{PLOS Comput. Biol.}
  \textbf{2009}, \emph{5}, e1000452.

\bibitem{us:orig}
G.~M. Torrie, J.~P. Valleau, \emph{Chem. Phys. Lett.} \textbf{1974}, \emph{28},
  578--581.

\bibitem{wham:1}
A.~M. Ferrenberg, R.~H. Swendsen, \emph{Phys. Rev. Lett.} \textbf{1989},
  \emph{63}, 1195--1198.

\bibitem{wham:2}
S.~Kumar, D.~Bouzida, R.~H. Swendsen, P.~A. Kollman, J.~M. Rosenberg, \emph{J.
  Comput. Chem.} \textbf{1992}, \emph{13}, 1011--1021.

\bibitem{dePablo:13}
M.~McGovern, J.~{de Pablo}, \emph{J. Chem. Phys.} \textbf{2013}, \emph{139},
  084102.

\bibitem{Laio:PRB:10}
F.~P. Y.~Crespo, F.~Marinelli, A.~Laio, \emph{Phys. Rev. B} \textbf{2010},
  \emph{81}, 055701.

\bibitem{Tiana:08}
G.~Tiana, \emph{Eur.~Phys.~J.~B} \textbf{2008}, \emph{63}, 235--238.

\bibitem{Bonomi:09}
M.~Bonomi, A.~Barducci, M.~Parrinello, \emph{J.~Comput.~Chem.} \textbf{2009},
  \emph{30}, 1615--1621.

\bibitem{Tiwary:14}
P.~Tiwary, M.~Parrinello, \emph{J. Phys. Chem. B} \textbf{2014}, \emph{119},
  736--742.

\bibitem{Filizola:12}
J.~M. Johnston, H.~Wang, D.~Provasi, M.~Filizola, \emph{PLoS Comp. Biol.}
  \textbf{2012}, \emph{8}, e1002649.

\bibitem{Kastner:11}
J.~K{\"a}stner, \emph{WIREs Comput.~Mol.~Sci.} \textbf{2011}, \emph{1},
  932--942.

\bibitem{US:ITS}
M.~Yang, L.~Yang, Y.~Gao, H.~Hu, \emph{J.~Chem.~Phys.} \textbf{2014},
  \emph{141}, 044108.

\bibitem{ensing05}
B.~Ensing, A.~Laio, M.~Parrinello, M.~L. Klein, \emph{J. Phys. Chem. B}
  \textbf{2005}, \emph{109}, 6676--6687.

\bibitem{Sagui:06}
V.~Babin, C.~Roland, T.~A. Darden, C.~Sagui, \emph{J.~Chem.~Phys.}
  \textbf{2006}, \emph{125}, 204909.

\bibitem{Guella:10}
E.~Autieri, M.~Sega, F.~Pederiva, G.~Guella, \emph{J.~Chem.~Phys.}
  \textbf{2010}, \emph{133}, 095104.

\bibitem{Voth:11}
Y.~Zhang, G.~A. Voth, \emph{J.~Chem.~Theory Comput.} \textbf{2011}, \emph{12},
  2277--2283.

\bibitem{Wacker_ramana}
V.~Imandi, N.~N. Nair, \emph{J. Phys. Chem. B} \textbf{2015}, \emph{119},
  11176--11183.

\bibitem{ravi_nair_NDM}
R.~Tripathi, N.~N. Nair, \emph{ACS Catal.} \textbf{2015}, \emph{5}, 2577--2586.

\bibitem{Branduardi:2012}
D.~Branduardi, G.~Bussi, M.~Parrinello, \emph{J. Chem. Theory Comput.}
  \textbf{2012}, \emph{8}, 2247--2254.

\bibitem{Hub:JCTC:10}
J.~S. Hub, B.~L. de~Groot, D.~van~der Spoel, \emph{J. Chem. Theory Comput.}
  \textbf{2010}, \emph{6}, 3713--3720.

\bibitem{cpmd1}
{Version~13.2}, \emph{{\tt CPMD} {P}rogram {P}ackage}, IBM Corp 1990-2011, MPI
  f{\"u}r Festk{\"o}rperforschung Stuttgart 1997-2001, \\see also {\tt
  http://www.cpmd.org}.

\bibitem{PBEGGA}
J.~P. Perdew, J.~A. Chevary, S.~H. Vosko, K.~A. Jackson, M.~R. Pederson, D.~J.
  Singh, C.~Fiolhais, \emph{Phys. Rev. B} \textbf{1992}, \emph{46}, 6671--6687.

\bibitem{Pseudopotential}
D.~Vanderbilt, \emph{Phys. Rev. B} \textbf{1990}, \emph{41}, 7892--7895.

\bibitem{car85}
R.~Car, M.~Parrinello, \emph{Phys. Rev. Lett.} \textbf{1985}, \emph{55},
  2471--2474.

\bibitem{nhc}
G.~J. Martyna, M.~L. Klein, M.~Tuckermann, \emph{J. Chem. Phys.} \textbf{1992},
  \emph{97}, 2635.

\bibitem{string_method}
W.~E, W.~Ren, E.~Vanden-Eijnden, \emph{Phys. Rev. B} \textbf{2002}, \emph{66},
  052301.

\bibitem{laio02a}
A.~Laio, J.~VandeVondele, U.~Rothlisberger, \emph{J. Chem. Phys.}
  \textbf{2002}, \emph{116}, 6941--6947.

\end{thebibliography}

\providecommand{\url}[1]{\texttt{#1}}
\providecommand{\urlprefix}{}
\providecommand{\foreignlanguage}[2]{#2}
\providecommand{\Capitalize}[1]{\uppercase{#1}}
\providecommand{\capitalize}[1]{\expandafter\Capitalize#1}
\providecommand{\bibliographycite}[1]{\cite{#1}}
\providecommand{\bbland}{and}
\providecommand{\bblchap}{chap.}
\providecommand{\bblchapter}{chapter}
\providecommand{\bbletal}{et~al.}
\providecommand{\bbleditors}{editors}
\providecommand{\bbleds}{eds.}
\providecommand{\bbleditor}{editor}
\providecommand{\bbled}{ed.}
\providecommand{\bbledition}{edition}
\providecommand{\bbledn}{ed.}
\providecommand{\bbleidp}{page}
\providecommand{\bbleidpp}{pages}
\providecommand{\bblerratum}{erratum}
\providecommand{\bblin}{in}
\providecommand{\bblmthesis}{Master's thesis}
\providecommand{\bblno}{no.}
\providecommand{\bblnumber}{number}
\providecommand{\bblof}{of}
\providecommand{\bblpage}{page}
\providecommand{\bblpages}{pages}
\providecommand{\bblp}{p}
\providecommand{\bblphdthesis}{Ph.D. thesis}
\providecommand{\bblpp}{pp}
\providecommand{\bbltechrep}{Tech. Rep.}
\providecommand{\bbltechreport}{Technical Report}
\providecommand{\bblvolume}{volume}
\providecommand{\bblvol}{Vol.}
\providecommand{\bbljan}{January}
\providecommand{\bblfeb}{February}
\providecommand{\bblmar}{March}
\providecommand{\bblapr}{April}
\providecommand{\bblmay}{May}
\providecommand{\bbljun}{June}
\providecommand{\bbljul}{July}
\providecommand{\bblaug}{August}
\providecommand{\bblsep}{September}
\providecommand{\bbloct}{October}
\providecommand{\bblnov}{November}
\providecommand{\bbldec}{December}
\providecommand{\bblfirst}{First}
\providecommand{\bblfirsto}{1st}
\providecommand{\bblsecond}{Second}
\providecommand{\bblsecondo}{2nd}
\providecommand{\bblthird}{Third}
\providecommand{\bblthirdo}{3rd}
\providecommand{\bblfourth}{Fourth}
\providecommand{\bblfourtho}{4th}
\providecommand{\bblfifth}{Fifth}
\providecommand{\bblfiftho}{5th}
\providecommand{\bblst}{st}
\providecommand{\bblnd}{nd}
\providecommand{\bblrd}{rd}
\providecommand{\bblth}{th}

\clearpage
\renewcommand{\arraystretch}{1.0}

\begin{table}[t]
\caption{Parameters for the two--dimensional model potential \label{table:uparams}}
\centering
\begin{tabular}{ l l l l l}
\hline
$i$ & $U_i^o$ (kcal~mol$^{-1}$)& $a_i$ (Bohr$^{-2}$) & $x_i^o$ (Bohr) & $y_i^o$ (Bohr)\\
\hline
1 & -17.885 & 139.2985 & 0.0000& 0.00000 \\
2 & -11.625 & 41.7895 & 0.3642 & 0.00000 \\
3 & -11.625 & 41.7895 & 0.4916 & 0.00000 \\
4 & -11.625 & 41.7895 & 0.6189 & 0.00000 \\
5 & -17.885 & 13.9298 & 0.1821 & 0.00000 \\
\hline
\end{tabular}
\end{table}
\clearpage

\begin{table}[t]
\caption{Free energy convergence in WS--MTD simulations for the double--well potential. The converged free energy barriers
from 86.7~ps of two--dimensional WT--MTD and the exact free energy values from the potential are also provided for comparison. 
\label{t:mtd:conv}}
\begin{center}
\begin{tabular}{| c | c | c | c | c | }
\hline
Method & $t_{\rm min}$~(ps) & $t_{\rm max}$~(ps) & \multicolumn{2}{|c|}{$\Delta F^\ddagger$ (kcal~mol$^{-1}$)}  \\ \cline{4-5}
       &            &                    &{\bf A}$\rightarrow$ {\bf B} & {\bf B}$\rightarrow$ {\bf A}  \\ \hline

WS--MTD&   0.0      &   0.6              &     10.5                    &       9.5                    \\
       &            &   1.4              &     10.0                    &       9.2       \\
       &            &   2.8              &     9.9                     &       9.3     \\ \cline{2-5}
       &   0.1      &   0.6              &     10.1                    &       8.8                    \\
       &            &   1.4              &     9.8                     &       9.2       \\
       &            &   2.8              &     9.8                     &       9.3     \\ \cline{2-5}
       & 0.2        &   1.4              &     9.7                     &       9.1                    \\
       &            &   2.8              &     9.8                     &       9.3                    \\ \cline{2-5}
       &  1.5       &   10.0             &     9.8                     &       9.3                    \\ \hline
WT--MTD &  -        &    -               &    9.8                      &       9.3     \\ \hline
Exact   &  -        &    -               &    9.8                      &       9.3      \\ \hline
\end{tabular}
\end{center}
\end{table}

\clearpage

\begin{table}[t]
\caption{Convergence in the computed forward free energy barriers for 
{\bf WA1}$\rightarrow${\bf WA2} 
with number of iterations.\label{table:wacker_walk}}
\centering
\begin{tabular}{ |c|c|}
\hline
Iteration & $\Delta F^\ddagger$ (kcal~mol$^{-1}$)  \\
\hline
1 &   19.4 \\
2 &   19.9 \\
3 &   20.1  \\
\hline
\end{tabular}
\end{table}
\clearpage

\clearpage

\clearpage

\clearpage

\clearpage

\clearpage

{\bf Table of Content Figure:} \\

\begin{center}
\includegraphics[height=3.5cm]{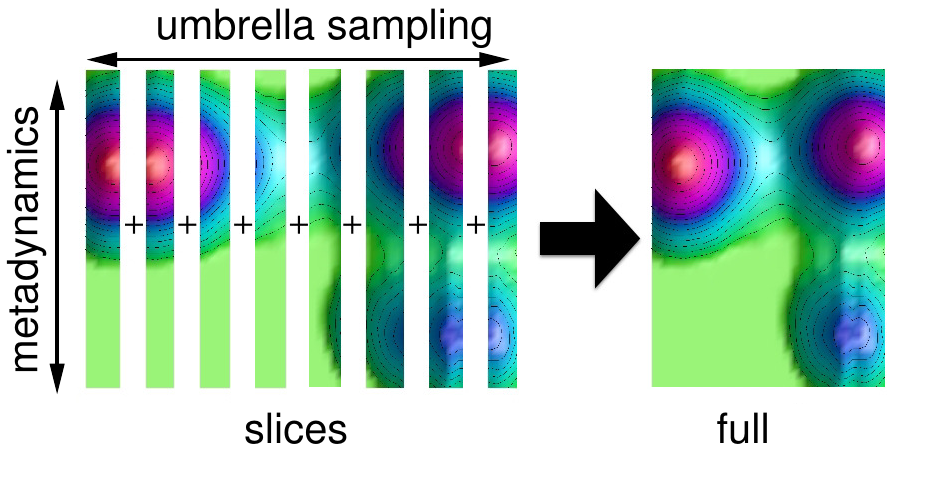}
\end{center}

{\bf Abstract for the Table of Content Figure:} \\

Here we present a technique to sample a high--dimensional free energy landscape 
as slices by combining umbrella sampling and metadynamics simulation sampling orthogonal
coordinates simultaneously. The full free energy surface is then reconstructed by combining these slices
using the Weighted Histogram Analysis Method.

\end{document}